# On Sink Mobility Trajectory in Clustering Routing Protocols in WSNs

N. Javaid[1], Q. Ain[1], M. A. Khan[1], A. Javaid[2], Z. A. Khan[3], U. Qasim[4]

*COMSATS Institute of Information Technology, [1]Islamabad, [2]Wah Cantt, Pakistan*

*[3]Faculty of Engineering, Dalhousie University, Halifax, Canada*

*[4]University of Alberta, Alberta, Canada*

Abstract—Energy efficient routing protocols are consistently cited as efficient solutions for Wireless Sensor Networks (WSNs) routing. The area of WSNs is one of the emerging and fast growing fields which brought low cost, low power and multi-functional sensor nodes. In this paper, we examine some protocols related to homogeneous and heterogeneous networks. To evaluate the efficiency of different clustering schemes, we compare five clustering routing protocols; Low Energy Adaptive Clustering Hierarchy (LEACH), Threshold Sensitive Energy Efficient Sensor Network (TEEN), Distributed Energy Efficient Clustering (DEEC) and two variants of TEEN which are Clustering and Multi-Hop Protocol in Threshold Sensitive Energy Efficient Sensor Network (CAMPTEEN) and Hierarchical Threshold Sensitive Energy Efficient Sensor Network (H-TEEN). The contribution of this paper is to introduce sink mobility to increase the network life time of hierarchal routing protocols. Two scenarios are discussed to compare the performances of routing protocols; in first scenario static sink is implanted and in later one mobile sink is used. We perform analytical simulations in MATLAB by using different performance metrics such as, number of alive nodes, number of dead nodes and throughput.

## 1. INTRODUCTION

WSNs are composed of sensor nodes which can sense, compute, store and transceive data of interest in environment. These sensor nodes are scattered in sensor field and capable of transmitting data to sink. In many applications, WSNs are used such as in military, environmental disaster areas. Some applications of WSNs are temperature monitoring, landslide detection; air pollution and structural monitoring. Nodes which are far away from sink consume more energy. Clustering technique is one of the popular mechanisms in which nodes select a Cluster Head (CH) for transmission. All nodes transmit their data to CH, where, it aggregates data and send to the Base Station (BS). Only few nodes are used to transmit at large distance so, less energy is consumed. The main idea of clustering is to reduce the network traffic from node to sink. Clustered protocols exhibit better performance in terms of energy consumption when compared to conventional protocols. Clustering sensor networks are classified in two types of network such as homogeneous network and heterogeneous network. In homogeneous network, all nodes have same energy level and in heterogeneous network, nodes have different energy levels. LEACH [1] and TEEN [2] are the examples of homogeneous protocols, while SEP [3] (Stable Election Probability) and DEEC [4] are heterogeneous Protocols. All clustering techniques consist of two phases; setup phase and steady state phase. In setup phase, formation of clusters and election of CHs is performed and in steady state phase, nodes transmit data to CH and it aggregates the data for sending to BS.

In earlier research, static sink is used to gather data in WSNs. The sensors near to sink consume more energy and die earlier in hierarchal protocols. The sink disconnects from network but rest of sensors has sufficient residual energy which is wasted. To overcome this problem, mobile sink is used which can move over the network to gather data. Sink mobility can be classified as uncontrollable and controllable mobility. Control able mobility is achieved by adding mobile intestinally in the network to carry sink node. Uncontrollable mobility is by attaching the sink node in network which is out of control.

In this paper, we compare five routing protocols in terms of their stability period, network life time and throughput. In case of stability period, DEEC performs better than other protocols because it is a heterogeneous protocol which has nodes with some extra energy known as advanced and super nodes. These nodes die later than normal nodes so can transmit more packets. When we consider network life time, H-TEEN [5] performs better than other protocols because, it is a threshold based protocol where nodes only transmit when they sense this value and consume less energy. The Contributions of this paper are: 1) The concept of sink mobility in clustering protocols, like, LEACH, TEEN, H-TEEN, CAMPTEEN, and DEEC. 2) Analyzing these protocols in order to compare the performance.

## 2. MOTIVATION

Research has shown that nodes near the sink deplete their battery power faster than the nodes apart due to heavy overhead of messages in hierarchal protocols. Sensors nearby sink are shared by more sensors to sink paths therefore consume more energy. This uneven energy depletion causes energy holes and leads to shorten network lifetime. Numerous researches have been conducted to mitigate this problem such as MobiRoute routing protocol and DT-MSM (delay tolerant mobile sink model). The position of sink is varied at different locations of the network field area from center to border axis in many routing schemes.

In LEACH, CH with large number of member nodes drains more energy than CH with smaller number of nodes. To mitigate this issue MLEACH protocol is proposed, it allows mobility of both sink and nodes. The CH selection criteria are based on remaining energy of nodes. It improves the network lifetime and throughput [6].

DEEC is a heterogeneous protocol having nodes with extra energy which can send packets to BS than other protocols. DEEC is compared with its variants to evaluate the performances of all proposed protocols [7].

MSWSN model uses mobile sink to gather data from static nodes. Model proposed to move sink with relative distance, direction and speed. It increases the delivery ratio, residual energy and network lifetime by one hop communication. Mobile sink transmits Buffer space to each sensor node for storing data. WCDMA technology is used to ensure energy efficient communication. Proposed model provides the relative random motion in sensor network and avoids threats in the multi-hop communication.

MobiRoute routing protocol proposed a model with a path predictable mobile sink to improve the packet deliver ratio. The sink is located at any point of network and the pause time of sink is longer than movement time. It has enough time to gather the data from nearby nodes. In this protocol all nodes should know the topological changes created due to sink mobility [8].

In [9] Rakhshan et al. also focused on maximizing the network lifetime of WSNs by using sink mobility. Energy formula that determines the moving times of sink based on residual energy of sensor nodes is proposed in this work. Furthermore, approach to calculate optimum path is used, which results in increasing the network lifetime.

In [10] Jafri et al. proposed mobility in multi-chain PEGASIS based protocol. In this work mobile sink is moved along its trajectory and stays for a sojourn time at sojourn location to guarantee complete data collection. This approach increased network lifetime to a distinct level, as compared to classical approaches.

In DT-MSM model, nodes can postpone the transmission until the sink stops at most favorable position. This model implements where the nodes can tolerate the delay in data delivery. As the sink location increases, optimal network life increases but there is delay in receiving data [11].

Chakrabarti et al. proposed a model for power saving in WSNs using predicable path of sink. The model is performed for networks only single hop communication is used and analyze the success in data collection, and quantify the power consumption of the network using queuing model [12].

This paper discusses a frame work to maximize the network lifetime in hierarchal protocol without delay in transmission of data packets by using a mobile sink. The sink mobility is used in both single hop and multi hop communication. We consider both homogeneous and heterogeneous networks and compare them by implementing the concept of sink mobility in routing protocols such as, LEACH, TEEN, DEEC and variants of TEEN. Their performances are observed in MATLAB. The position of sink is varied at different locations of the network field area from center to border axis in many routing schemes. We implement sink on the top axis of the network region and compare the results of all routing protocols.

## 3. ENERGY EFFICIENT ROUTING PROTOCOLS

Clustered sensor networks are categorized in two types of network; homogeneous and heterogeneous networks. In homogeneous network, all nodes are equipped with same energy, there is no predetermined cluster heads that control the cluster. Although this may be costly because each node is designed to be CH, but still reliable due to its independence of nominating the CHs. It is evident that in long transmission of CH to sink, if the clustered nodes are same in all rounds, they will be over-loaded with data and will expire before other nodes. One way to ensure this is to rotate the CH randomly as proposed in LEACH. In this way, chance of being CH is provided to all nodes and they will die at same time.

In heterogeneous network, there are many types of nodes other than normal nodes such as advanced and super with different energy level and functionality. The use of different energy is that it reduces the hardware cost of other network. Some extra energy is given to nodes other than normal nodes so, they have more chances to be CH than normal nodes. CHs are predetermined in heterogeneous protocol that controls the cluster.

Ideally, sensor networks should perform its functionality as long as possible. Conventional routing schemes are not feasible in terms of energy constrained. Energy efficient scheme is used to improve the routing protocols. There are some energy efficient routing protocols which are discussed below related to homogeneous and heterogeneous networks.

### A. LEACH

LEACH is a first clustering based protocol which provides same initial energy among all nodes. After cluster formation, nodes select one CH which requires minimum communication energy. LEACH uses random election of CH because if CHs are fixed throughout the system they will die quickly and network disconnects. It aggregates the data from member nodes for transmitting to BS which decreases the energy consumption and enhance lifetime of network. LEACH can achieve factor of 8 in decreasing energy consumption than conventional protocols. Direct transmissions and Minimum transmission energy (MTE) are the conventional protocols. In direct communication protocol, nodes are directly connected to BS means they send data without any other sensor node. The nodes which are far away from BS die earlier because they require large amount of transmission energy due to large distance from BS. This protocol is acceptable only if BS is close to the nodes. In MTE, nodes send data to BS through intermediate nodes. These nodes provide a routing path for other nodes and also sense the environment. The intermediate nodes are selected by using transmit amplifier energy. If there are three nodes A, B and C in network, the transmission is done by node A to C through B if and only if this condition fulfills [1].

$$E_{Tx}(k, d = d_{AB}) + E_{Tx}(k, d = d_{BC}) < E_{Tx}(k, d = d_{AC}) \tag{1}$$

The transmission energy from A to B and from B to C is less than the transmission energy of nodes A to C then the node B will be elected as intermediate node. In MTE the nodes which are nearby to BS will die more quickly than the nodes which are far away. Nearby nodes are used as intermediate nodes which consume more energy than other nodes. In LEACH, radio interference is produced when multiple clusters transmit at a same time. To reduce this interference, each CH uses CDMA code and transmits data to its members. They transmit data in an order which reduces the interference.

### B. TEEN

TEEN is a clustering routing protocol and related to reactive network. It is an extension of LEACH protocol using same clustering scheme with an extra entity which is a threshold value. TEEN protocol is data centric in which nodes get data on certain parameters such as temperature, pressure, velocity and humidity e.g. if the node sense temperature greater than 50 then it sends data to BS otherwise remain in sleep mode. If adjacent nodes have similar data then it is more favorable to aggregate the data from each node rather than every node sends it separately. The main features of TEEN protocol are that nodes have to transmit to their CH to dissipate less energy, additional computation is done only by CH to save energy, CHs present at high level of hierarchy have to transmit data which consume more energy. To overcome this problem, all nodes will be CH for a time period T (cluster Period). In TEEN, nodes sense environment all the time and transmission is done only when there is a drastic change. Due to minimal transmission less energy is consumed. CHs broadcast two threshold values to its members which are,

- Hard threshold
- Soft threshold

Hard threshold is the absolute value, when a node sense this value transmitter is switched on and send data to the CH. A small change in the sensed value is soft threshold. Node sense the environment, when hard threshold is achieved node transmits the sensed data. Hard threshold decreases the transmission of data because nodes only transmit when they sense the drastic change. Soft threshold also reduces the number of transmission by ignoring small changes in sensed value.

TEEN protocol is applicable in time critical data sensing applications, where user needs the data instantaneously. Data transmission consumes more energy so; nodes only sense the environment and transmit only when they sense a threshold value. The drawback of TEEN protocol is that nodes will never communicate if the threshold value is not reached, the notification will not receive by user even if all nodes die. It is implemented where no collision exists. TDMA scheduling can be used to avoid this problem but it introduces the delay. This scheme is not suitable where user needs the data in periodic way.

CH selection criteria are same in both TEEN and LEACH protocol. The decision is made by a random number selection between 0 and 1, member nodes will be CH if selected number is less than the given threshold equation [1],

$$T(n) = \begin{cases} \frac{p}{1-p(r mod \frac{1}{p})} & \text{if } n \epsilon G \\ 0 & \text{otherwise} \end{cases} \qquad (2)$$

Where P is the percentage of CHs, r is the current round and G is the set of nodes which are eligible to b CHs. 1p are the $e_{pochs}$ rounds after which a node is eligible to b a CH. When it is selected, data is transmitted to BS in compressed form.

C. DEEC

DEEC is designed for heterogeneous network in which some nodes are advanced with more energy than normal nodes. All the nodes are equipped with different energy levels to increase the network lifetime. All nodes have same energy in initial level, by reenergizing the sensor nodes either by adding nodes which have high energy or to provide energy to already existing nodes. The energy nodes having low energy will die earlier than other nodes. CH selection is based on the remaining energy of node and average energy of the network. The data is gathered by CHs from its node members and forward to the BS. $e_{pochs}$ is different for each node according to its residual and initial energy. The nodes having high residual and high initial energy will be CHs more time than with low energy nodes. DEEC is used to increase the network life time. In DEEC, there are two types of nodes such as advanced and normal nodes but DEEC is multi level heterogeneous network. The total initial energy of this network is given by,

$$E_{total} = \sum_{i=1}^{N} E_0(1+a_i) = E_0(N + \sum_{i=1}^{N} a_i) \qquad (3)$$

In DEEC, there are some nodes with extra energy represented by a and m is the probability of extra nodes. CH selection is based on the ratio between residual energy of each node and average energy of the network. $e_{pochs}$ of being cluster head is different according to their initial and residual energy. The nodes having high initial and residual energy will have more chances to become cluster head (4). CH selection is based on the threshold equation (4),

$$p_i = p_{opt}[1 - \frac{\bar{E}(r) - E_i(r)}{\bar{E}(r)}] = p_{opt} \frac{E_i(r)}{\bar{E}(r)} \qquad (4)$$

Where $p_{opt}$ is the reference value of the average probability of $p_i$, $\bar{E}(r)$ is the average energy of the network and $E_i(r)$ is the residual energy of the network at round r. Total number of CHs per round per $e_{poch}$ is

$$\sum_{i=1}^{N} p_i = \sum_{i=1}^{N} p_{opt} \frac{E_i(r)}{\bar{E}(r)} = p_{opt} \sum_{i=1}^{N} \frac{E_i(r)}{\bar{E}(r)} = N p_{opt} \qquad (5)$$

$$T(s_i) = \begin{cases} \frac{p_i}{1 - p_i(r \bmod \frac{1}{P_i})} & if \ s_i \epsilon G \\ 0 & otherwise \end{cases} \qquad (6)$$

Threshold value is given in above equation on which CH election criteria is based. In two level heterogeneous network the value of $p_{opt}$ is given by,

$$p_{adv} = \frac{p_{opt}}{1 + am}, p_{nrm} = \frac{p_{opt}(1 + a)}{(1 + am)} \qquad (7)$$

Then use the above $p_{adv}$ and $p_{nrm}$ instead of $p_{opt}$ in above equation for two level heterogeneous network as supposed in equation (3.8)

$$p_i = \begin{cases} \frac{p_{opt} E_i(r)}{(1+am)\bar{E}(r)} & if \ s_i \ is \ the \ normal \ node \\ \frac{p_{opt}(1+a) E_i(r)}{(1+am)\bar{E}(r)} & if \ s_i \ is \ the \ advanced \ node \end{cases} \qquad (8)$$

It can be extended to multi-level heterogeneous network which is given as;

$$p_{multi} = \frac{p_{opt} N(1 + a_i)}{(N + \sum_{i=1}^{N} a_i)} \qquad (9)$$

Above $p_{multi}$ is used instead of $p_{opt}$ to get pi for heterogeneous node. pi for the multilevel heterogeneous network is given as;4

$$p_i = \frac{p_{opt} N(1 + a) E_i(r)}{(N + \sum_{i=1}^{N} a_i)\bar{E}(r)} \qquad (10)$$

In DEEC we estimate average energy $E(r)$ of the network for any round r as in;

$$\bar{E}(r) = \frac{1}{N} E_{total}(1 - \frac{r}{R}) \qquad (11)$$

$R$ denotes total number of rounds and is estimated as follows;

$$R = \frac{E_{total}}{E_{round}} \qquad (12)$$

If the residual energy of the nodes is higher than average energy of network than it have more chances to become CH.

D. H-TEEN

H-TEEN is a variant of TEEN protocol, introducing a hierarchy of clustering to better cope with large network area. When number of layers in hierarchy is small TEEN consumes lot of energy because of larger distance so, H-TEEN performs better due to less consumption of energy in large network. H-TEEN is a 4 layer hierarchal clustering where sensors self-organize into clusters and build a tree of transmissions and propagate data to the CH. CH selection is same as in TEEN and LEACH. The threshold equation is given as (5),

$$T(n) = \begin{cases} \frac{p_c}{1-p_c(r mod \frac{1}{p_c})} & \text{if } n \epsilon G \\ 0 & \text{otherwise} \end{cases} \qquad (13)$$

Here r is the current round and G is the set of nodes eligible to become CH. The threshold value increases as rounds pass; alive nodes become a CH after rounds 1p. When a node becomes it broadcasts an advertisement message to all member nodes. Nodes receive the message and elect the CH which depends on signal strength. When node decides cluster, it transmits a message to CH belonging to that cluster. This process is done by using CSMA MAC protocol to avoid collisions. The CH broadcasts a TDMA schedule for transmissions to all its members. Now the next level of hierarchy is build. Previous CHs decides whether they could be CH for next level of hierarchy. If they are eligible then sends an advertisement message otherwise another node is selected to be CH. After clustering, data transmission starts from nodes to CH. TDMA schedule is broadcasted to all member nodes for transmission. To reduce interference CDMA code is used, each node chooses a different code. Moreover, two threshold values are used, the hard and soft threshold values. Nodes sense the field continuously, when hard threshold is reached node transmit data to BS. Soft threshold is small change in that sensed value.

E. CAMP-TEEN

CAMP-TEEN is the extension of TEEN protocol, most suitable for the application of land slide prediction. Nodes sense the slight movement of soil and change in parameters that occur before land slide. CAMP enhances localization and energy efficiency of multi-hop routing protocol and TEEN is an extended version of LEACH which saves energy by using threshold values. It is useful in landslide prediction applications because each rock have different threshold values. In CAMP-TEEN, one node broadcasts a beacon pulse. Nodes which are nearby to that node, receive this beacon and sends an acknowledgement return to beacon node. The acknowledgment has the distance between nodes and beacon node based on RSSI (Received signal strength indication). It constructs the neighborhood table for each node until all nodes have their neighboring table. CAMP uses distributed clustering in which CH is selected on the basis of local information of nodes. In CAMP-TEEN, CH selection criteria depends on a timer which is given as [6],

$$T(v) = \frac{K}{E} - \alpha \qquad (14)$$

Where K is the proportionality constant which is taken as 1, E is the normalized energy of the node and α is the random number between 0 and 1. Timer starts for every node by using above equation. The node with least timer value will have high energy as they are inversely proportional to each other. The high energy node will be elected as a CH then the nodes in neighboring of CH will terminate their timers. CHs broadcast TDMA schedule to their cluster members. Nodes transmit data to CH, it collects the data and forwards it to BS. In table (1), some features of protocols are described in terms of their types of communication, network and routing. Moreover, their CH selection criteria are also mentioned.

4. RADIO DISSIPATION MODEL

According to radio energy dissipation model, the energy expended by transmitting L bit message over distance d is given as, [4]

$$E_{Tx}(L,d) = \begin{cases} L.E_{elec} + L.\epsilon_{fs}.d^2 & \text{if } d < d_0 \\ L.E_{elec} + L.\epsilon_{mp}.d^4 & \text{if } d > d_0 \end{cases} \quad (15)$$

Where $E_{fs}$ is energy dissipated to run the transmitter or receiver circuit, $E_{fs}$ is free space transmit amplifier if $d_{max}toBS < d_0$, $E_{mp}$ is multi path transmit amplifier if $d_{max}toBS > d_0$ and $d$ is the distance between sender and receiver. Total energy dissipated during a round is given as [4],

$$E_{round} = L(2NE_{elec} + NE_{DA} + k\epsilon_{mp}d_{toBS}^4 + N\epsilon_{fs}d_{toCH}^2) \quad (16)$$

Where k is the number of clusters, $d_{toBS}$ is the distance between the CH and BS and $d_{toCH}$ is the distance between cluster members and CH[4].

$$d_{toCH} = \frac{M}{\sqrt{2\Pi k}}, \quad d_{toBS} = 0.765\frac{M}{2} \quad (17)$$

$$k = \frac{\sqrt{N}}{\sqrt{2\Pi}}\sqrt{\frac{\epsilon_{fs}}{\epsilon_{mp}}}\frac{M}{d_{toBS}^2} \quad (18)$$

TABLE I
Features of DEEC, LEACH, TEEN and its VARIANTS

| Protocols | Network Type | Communication | Routing Type | CH Selection Criteria |
|---|---|---|---|---|
| DEEC | Heterogeneous | Single-hop | Proactive | Average energy of network and residual energy of node |
| LEACH | Homogeneous | Single-hop | Proactive | Threshold based probability |
| TEEN | Homogeneous | Multi-hop | Reactive | Threshold based probability |
| H-TEEN | Homogeneous | Multi-hop | Reactive | Threshold based probability |
| CAMP-TEEN | Homogeneous | Single-hop | Reactive | Timer based probability |

5. SINK MOBILITY

The energy of nodes near to sink exhausted very quickly in hierarchal protocols where the base station is fixed, as a result networks get disconnected. To overcome this problem and prolong the life time of network mobile sink is used to collect the data. Sink changes its position randomly according to the requirement of mobility. There are many ways in which sink can move in a network of, it can move on the top of network, bottom of network, diagonally in network and in many other directions. Before the sink changes its position, it stops for a fixed amount of time to collect the data from sensors within its range called pause time. During pause time, the sink broadcasts a beacon frame to its neighboring node for transmitting the data packets. When node sends the data, sink broadcasts another beacon frame to stop transmission which reduces the packet drop. The network life time can be extended if mobile sink balances the traffic load of nodes. To minimize the traffic load shortest path routing is used.

6. SIMULATION RESULTS

To evaluate the performance of different homogeneous and heterogeneous protocols, we have implemented it in MATLAB. The simulation has been performed on a network of 100 nodes and a base station (fixed and mobile). The nodes are placed randomly in the network. The field size is 100m x100m and number of rounds is 5000 in all scenarios. Our goal is to compare the performance of DEEC, LEACH, TEEN and two extensions of TEEN protocol which are H-TEEN and CAMPTEEN. Here we use three different metrics to analyze and compare the performance of the protocols which are, number of alive nodes, number of dead nodes, throughput (total rate of the data sent over the network) in two different scenarios as,

- Static Sink
- Mobile Sink

TABLE I
Parameter Values

| Parameters | Values |
|---|---|
| Network field | 100 m,100 m |
| Number of nodes | 100 |
| $E_0$ (initial energy) | 0.5J |
| Data bit | 4000 |
| $E_{elec}$ | 50 nJ/bit |
| $E_{fs}$ | 10 nJ/bit/ $m^4$ |
| $E_{mp}$ | 0.013/pJ/bit/ $m^4$ |
| $E_{DA}$ | 5 nj/bit |
| $P_{opt}$ | 0.1 |

For the sake of consistency, we have taken same energy model parameter for all the protocols which are given in table1. Fig. 1 shows network life time of all routing protocols in first scenario when the sink is fixed and deployed in center of the network. We observe that by comparing LEACH, TEEN and DEEC the stability period of LEACH is shorter almost 50% and 55% less than DEEC and TEEN because in LEACH, energy of all nodes are same it takes no advantage of nodes that have more energy than other nodes and having no threshold values. The network life time of TEEN is greater than LEACH and DEEC because in TEEN hard threshold and soft threshold values are used for nodes to transmit the data. Nodes are in sleep mode in TEEN, they only transmit when sensed the threshold value so, data transmission is done less frequently and less energy is consumed. DEEC generates un-even number of CHs for every round that can disturb performance of network where optimal number of CHs is necessary to enhance network life which is implemented in TEEN. When we consider H-TEEN the nodes remain alive for maximum number of rounds more than 5000 because it is a four level hierarchal protocol. CHs at increasing level have to transmit data to base station. When the number of layers used in hierarchy increases the transmission become shorter and less energy is consumed. TEEN is two level hierarchy protocol and H-TEEN is four level hierarchy protocol so it performs better than TEEN and other protocols in terms of network life time. Fig. 3 shows the throughput of all protocols. The throughput of DEEC is significantly larger than that of TEEN, LEACH, H-TEEN and CAMP-TEEN. In DEEC, nodes can transmit data continuously to base station while in TEEN and its variants, there is limited transmission because they are threshold based protocol and have limited information to share with sink. Another reason is that DEEC is a multilevel heterogeneous protocol having some advanced and super nodes which have extra energy than other normal nodes so they will die later than normal nodes and these nodes could be CHs more times than normal nodes. We compare all the protocols in second scenario where the base station is mobile on the top of the network. Each CH sends its data to, when sink moves toward clustered nodes having minimum distance between them. Fig. 4 and Fig. 5 shows the network life time of all routing protocols.

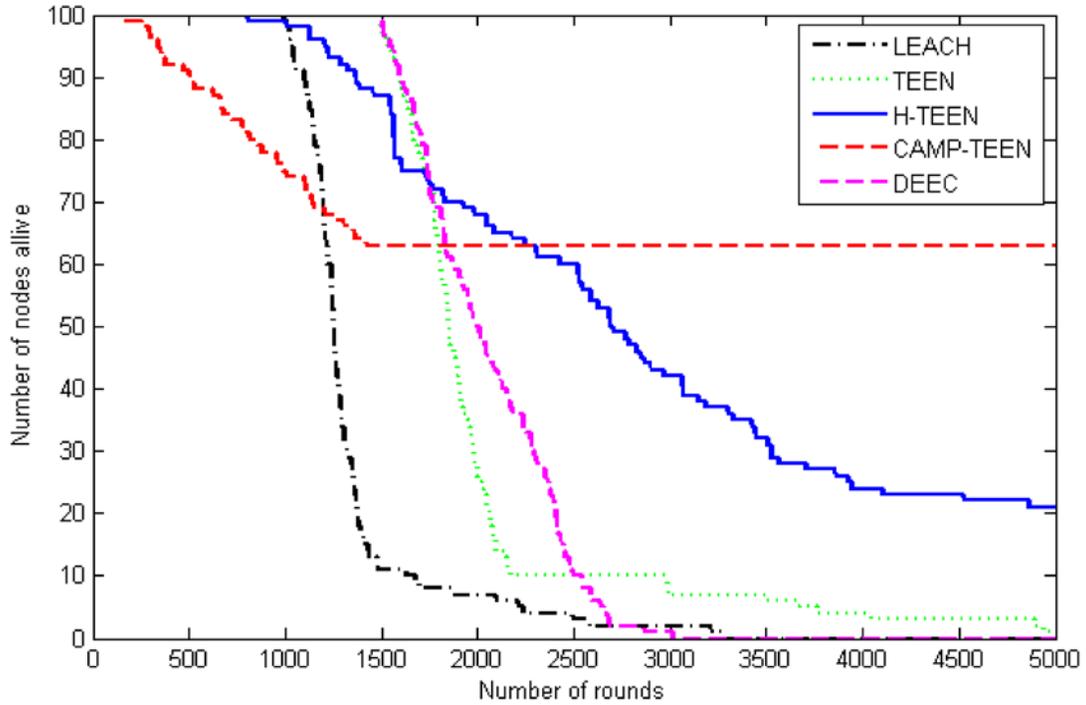

Fig. 1. Nodes alive during rounds

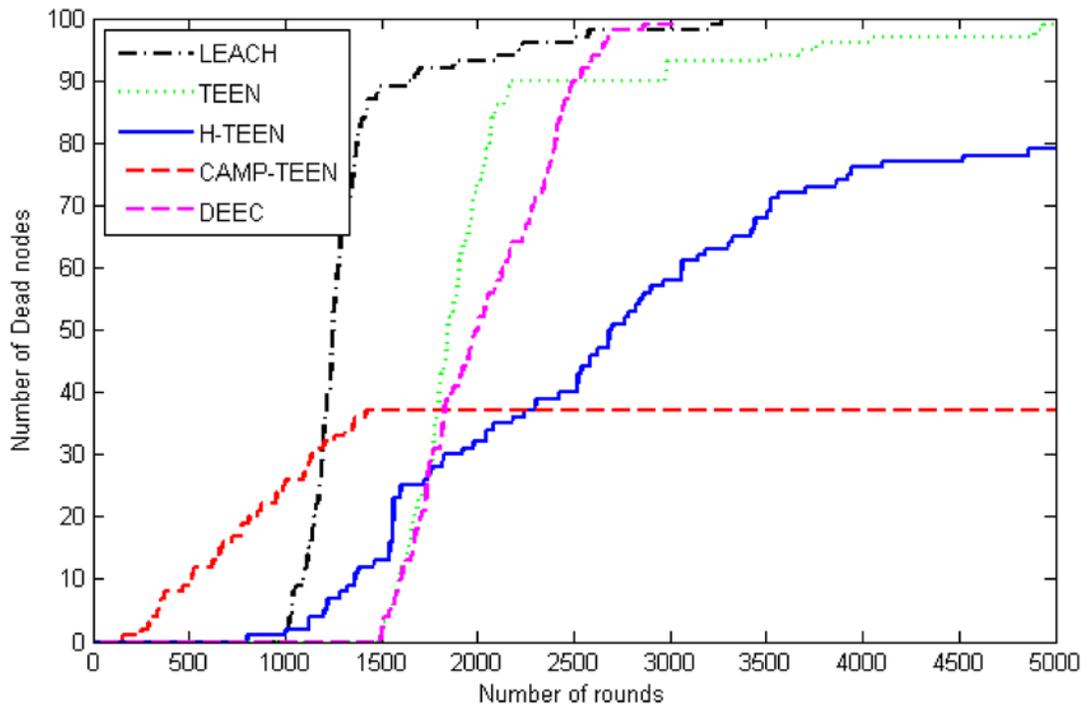

Fig. 2. Nodes dead during rounds

We examine that due to mobility of sink, stability period of protocols decreases because when the node is in the center of network it has equal distance to all nodes, they consume same amount of energy. Now sink is on top of network, the nodes which are at greater distance from sink will die quickly. Nodes consume greater energy at greater distance so die earlier. The network life time increases in only Mob-HTEEN because it enables to achieve less traffic load to nodes and reduces the delay in transmission.

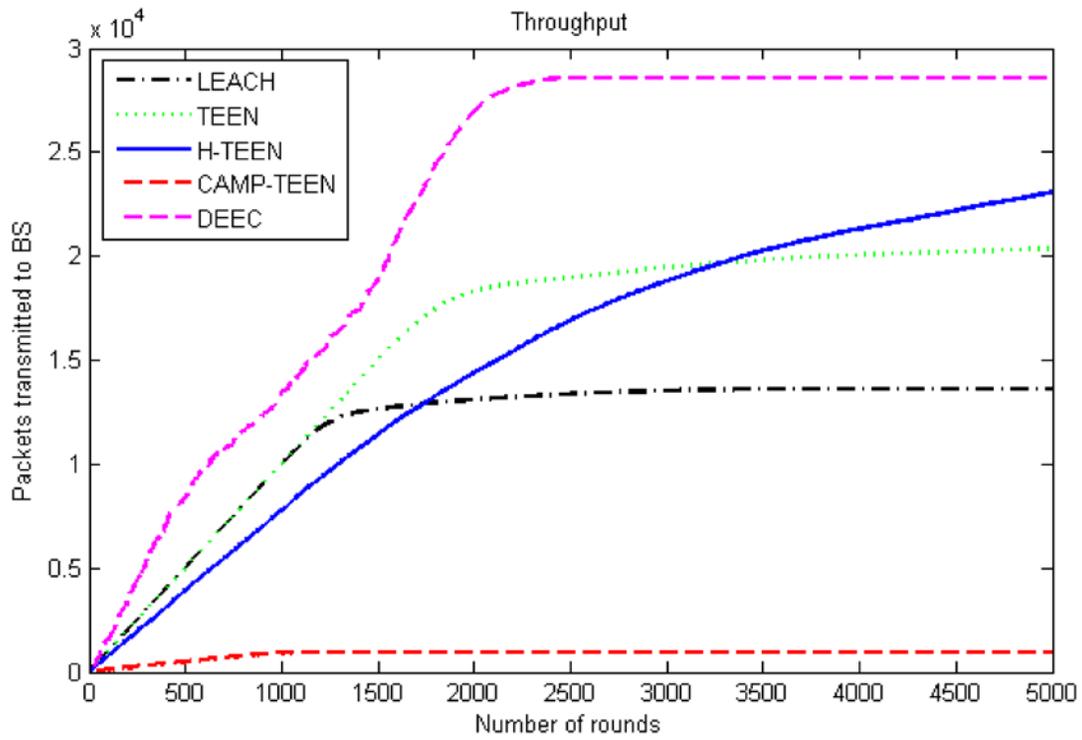

Fig. 3. Packets sent to BS

When sink moves, CHs near to sink transmits the data so consumed less energy and maximize the network lifetime.

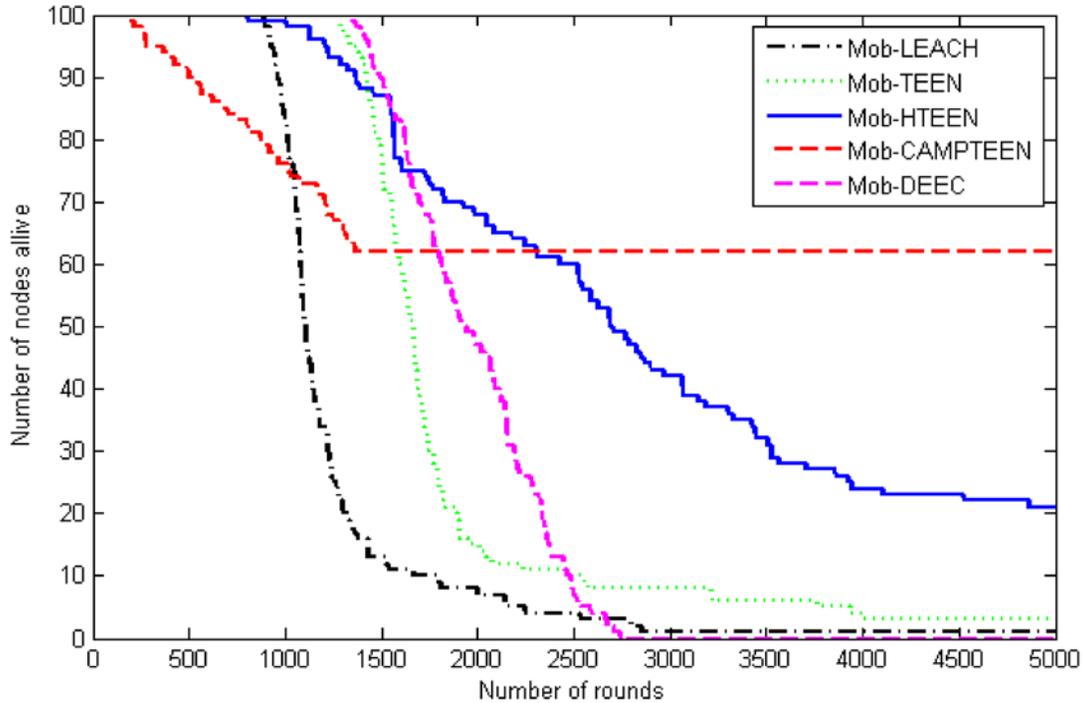

Fig. 4. Nodes alive during rounds

Fig. 6 shows the throughput of all proposed protocols. Less energy is consumed in Mob-HTEEN because clustered node only transmits when there is minimum distance between CHs and sink. When the CHs are in range of sink they transmit data directly. Due to minimum distance less energy is consumed than in fixed base station. Other protocols perform better in static sink case because CHs have same almost same distance to sink. In this case sink is away from nodes so, the CH present at large distance from sink consume more energy which decreases network lifetime. Throughput of Mob-DEEC is again better than all other protocols in this scenario as in case of fixed base station.

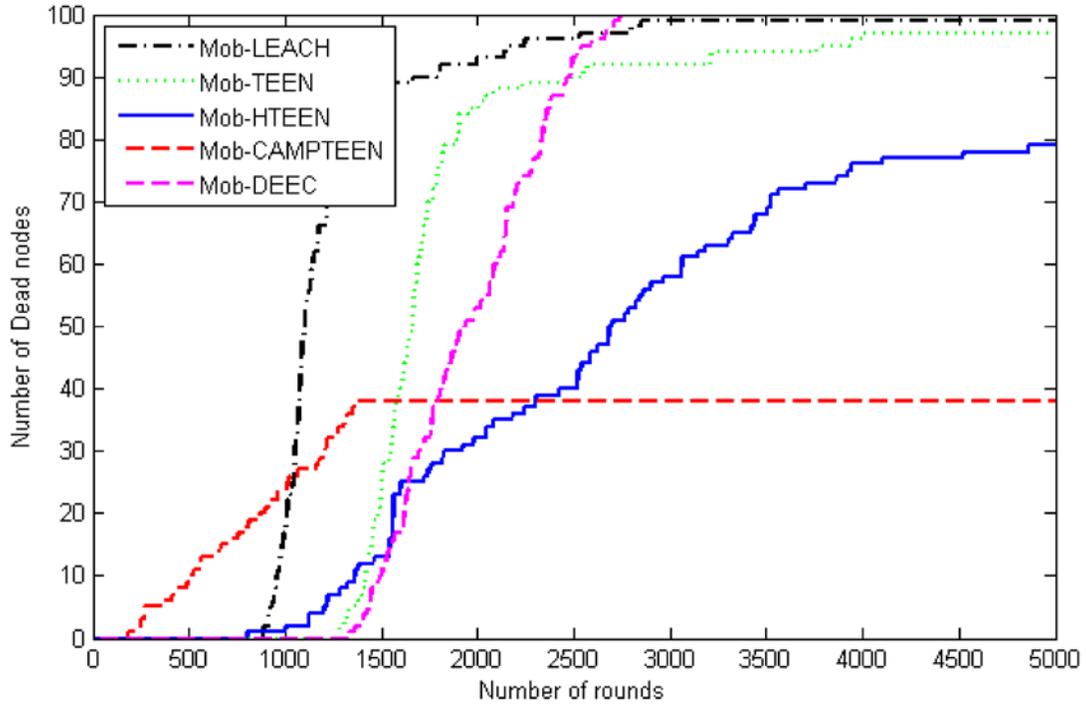

Fig. 5. Nodes dead during rounds

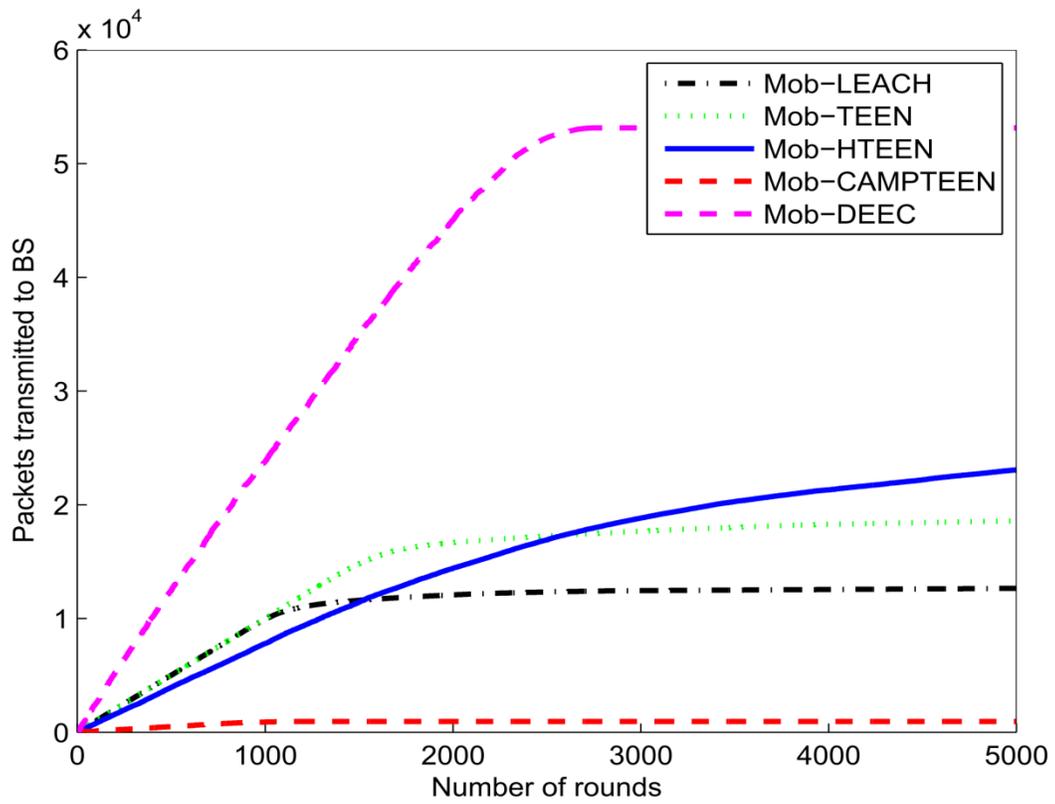

Fig. 6. Packets sent to BS

Firstly we compare all the protocols with static sink such as LEACH, DEEC, TEEN and its variants. Then we compare these protocols in by using mobile sink. We examine that results with mobile base station is better due to increasing network life time and throughput. DEEC is more scalable than other protocols and consume less energy. H-TEEN is highly energy efficient due to its hierarchy. Only HTEEN performs better in mobile sink and all other protocols give good results in static sink. Fig. 7 and Fig. 8 shows the bar plot of throughput and alive nodes of five protocols. The average values of all protocols are found in terms of throughput and number of alive nodes. The average throughput of LEACH, TEEN, HTEEN, CAMPTEEN and DEEC in case of static sink is 11574, 15745, 14889, 853, and 40872. In case of mobile sink the average throughput of these protocols are 10771, 14582, 16546, 948, and 22552. The throughput of DEEC in both cases is better than other protocols because of heterogeneous nature. By comparing network lifetime of all protocols in both cases such as mobile sink and static sink it is concluded that network lifetime and throughput of HTEEN is increased in case of mobile sink. Other protocols are performing better in static sink case. The average number of alive nodes of LEACH, TEEN, HTEEN, CAMPTEEN, and DEEC are 27, 41, 59, 69and 41 in case of static sink. In case of mobile sink, the average number of alive nodes are 25, 37, 68, 68, and 39. The average number of alive nodes is greater in TEEN and its variants because they are hierarchal clustered protocols and have threshold values which decrease the energy consumption.

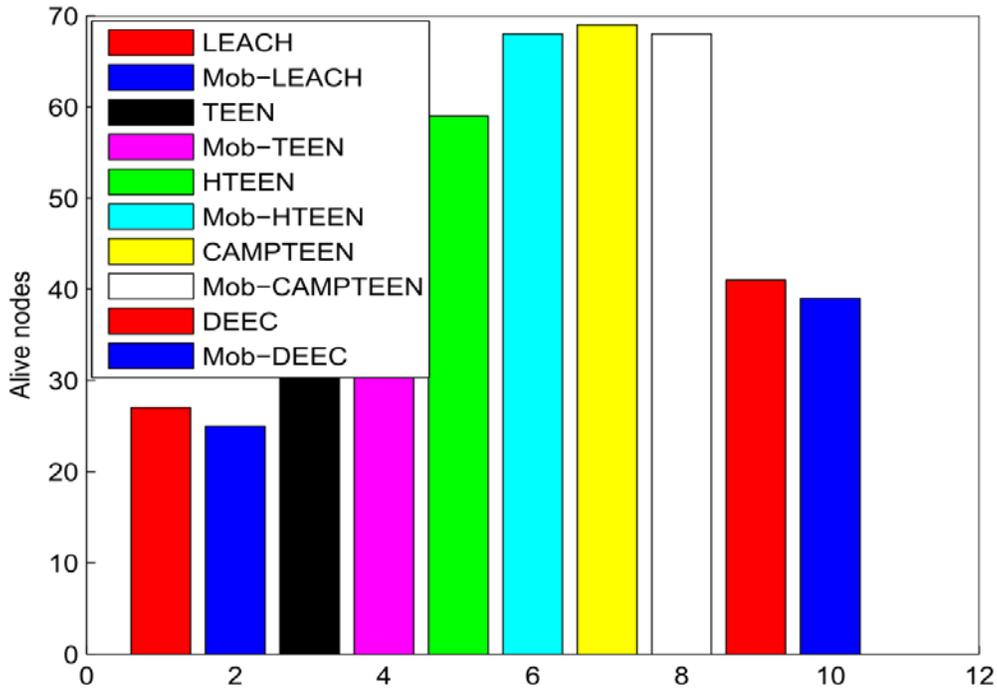

Fig.7. Average number of alive nodes

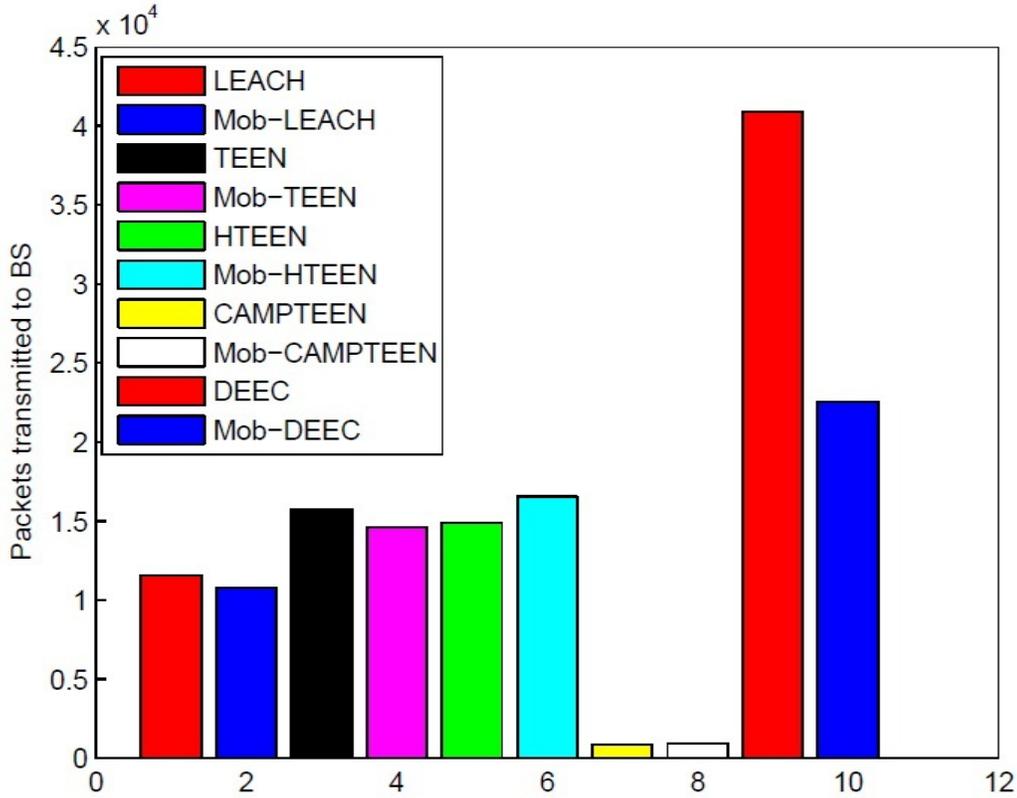

Fig. 8. Average throughput of alive nodes, number of dead nodes and throughput.

1. CONCLUSIONS

Energy efficiency and network lifetime are challenging issues in WSNs. Clustering technique have been proposed to resolve these issues by using different clustering schemes. We conclude from our analytical simulations that DEEC performs better in sending packets among all protocols. H-TEEN is more energy efficient because of hierarchical clustering and threshold value. Then, we introduce the mobility of sink in all proposed protocols to compare their performances. HTEEN outperforms in case of mobile sink and all other protocols perform better in case of static sink.